\documentclass[showpacs,preprintnumbers,amsmath,amssymb]{revtex4-1}

\usepackage{epsfig}
\usepackage{epsfig,amsmath}
\usepackage{graphicx}
\usepackage{dcolumn}

\usepackage{stmaryrd}
\usepackage{mathrsfs}
\usepackage{pifont}
\usepackage{amsthm}
\usepackage{amssymb}
\usepackage{amsmath}
\usepackage{epsfig,amsmath}
\usepackage{graphicx}
\usepackage{dcolumn}
\usepackage{bm}

\newcommand{\beq}{\begin{equation}}
\newcommand{\eeq}{\end{equation}}
\newcommand{\beqa}{\begin{eqnarray}}
\newcommand{\eeqa}{\end{eqnarray}}

\newcommand{\da}{\dagger} 
 \newcommand{\si}{\sigma}

\newcommand{\non}{\nonumber}

\def\pra#1{{ Phys.\ Rev. A\/} {\bf#1}}
\def\prb#1{{ Phys.\ Rev. B\/} {\bf#1}}

\def\prl#1{{ Phys.\ Rev.\ Lett.} {\bf#1}}

\begin{document}

\title{Perturbation Methods for Non-Markovian Quantum State Diffusion Equation}

\author{Jie Xu$^{1}$}
\email{Email address: jxu2@stevens.edu}
\author{Xinyu Zhao$^{1}$}
\author{Jun Jing$^{1,3,4}$}
\author{Lian-Ao Wu$^2$}
\author{Ting Yu$^1$}
\email{Email address: ting.yu@stevens.edu}

\affiliation{$^1$Center for Controlled Quantum Systems, and the
Department of Physics and Engineering Physics, Stevens Institute of
Technology, Hoboken, New Jersey 07030, USA \\  $^2$ Department of Theoretical
Physics and History of Science, The Basque Country University (EHU/UPV), PO Box 644, 48080 Bilbao, Spain\\ $^3$ Institute of Atomic and Molecular Physics, Jilin University, Changchun 130012, China\\ $^4$ Jilin Provincial Key Laboratory of Applied Atomic and Molecular Spectroscopy (Jilin University), Changchun 130012, China}

\date{\today}

\begin{abstract}
Two perturbation methods for the non-Markovian quantum state diffusion (NMQSD) equation are investigated in this paper. The first perturbation method under investigation is based on a functional expansion of the NMQSD equation, while the second expands the NMQSD equation in terms of coupling strength between the system and its environment. We compare the two perturbation methods by solving the dynamics of a bipartite system in which the two perturbation methods can be compared with the exact NMQSD equation. Additionally, as an application, we provide an analytical solution for a special family of system's initial states, and discuss the entanglement dynamics based on this solution.
\end{abstract}


\maketitle

\section{Introduction}

The Di\'{o}si-Gisin-Strunz non-Markovian quantum state diffusion (NMQSD) equation has provided a powerful tool in solving the dynamics of open quantum systems coupled to a non-Markovian bosonic environment \cite{Diosi}.  For example, when the system-environment coupling is strong or the environment surrounding the system of interest is structured,  the non-Markovian dynamics of the system cannot be accurately described by a traditional Lindblad Markov master equation  \cite{Nielson,Breuer,GardinerZoller,PLKnight,HPZ92,Goan07,Yu-Eberly06,Anas-HU,WMZhang}.  It has been shown that the Di\'{o}si-Gisin-Strunz NMQSD equation can be applied to solving non-Markovian dynamics for finite dimensional
quantum systems,  such as a multi-spin system or an $N$-level system  \cite{Xinyu,Jing-Yu12,Jun-Yu10,Jun-Wu-Yu12}, and continuous
variable systems, such as quantum Brownian motion, coupled cavities, and optomechanical oscillators  \cite{Strunz-Yu04,Strunz-Yu99,Strunz01}. Recently,  the NMQSD equation has been extended to the systems coupled to fermionic environments by introducing a Grassmann type of noise \cite{XinyuF, Wufu,Chen-You}.

Solving an NMQSD equation in realistic situations can be a challenging work. Therefore, a non-Markovian perturbation method based on the NMQSD equation is essential for practical applications and numerical implementations.
Several perturbative methods such as the post-Markov approximation, the weak-coupling approximation, and the functional perturbation are proposed  \cite{Strunz-Yu04,Yu-Strunz99,Gaspard-Nagaoka} to tackle practical physics models including photonic band-gap material \cite{Photon}, quantum dots \cite{Qdot}, chemical and biological systems \cite{Alex1,Alex2,Brumer}. It is interesting to note that a new method using hierarchy of stochastic equations is proposed in \cite{Sub-Strunz}, and an efficient numerical solution to the spin-boson model with functional expansion is presented in \cite{You-Pert}. However,  the non-Markovian perturbation approach is still under-developed due to the complexity arising from interplays between the system-environment coupling and the intrinsic environmental structure. From many aspects, the precise physical meanings underlying the perturbative approach are still unclear. For example, in the case of functional expansion, we still do not know the contribution of each term to
the non-Markovian dynamics for the problem under consideration \cite{Yu-Strunz99}.

The main purpose of this paper is to investigate the perturbation approaches to the NMQSD equation  that can be cast into a time-local form in a non-Markovian regime. In the case of long environmental correlation time and strong coupling between the system and the environment, we consider the functional expansion for the NMQSD equation. The NMQSD equation and the corresponding master equation derived from such an approximation can greatly reduce the difficulties arising from numerical simulations. The second
perturbation approach to be discussed in this paper is based on the expansion of the coupling
strength between the system and its environment.  If the coupling strength is weak, and the
quantum memory time of the environment is long, we show that the NMQSD equation derived from the weak-coupling assumption can be a good approximation in describing
the dynamics of system in a non-Markovian regime. An exactly solvable model is chosen as an example \cite{Xinyu} to discuss the both perturbation approaches.
As an illustration to the perturbation approaches discussed in this paper, the dynamics of entanglement under the approximate QSD equations is discussed. Given the important  applications of entanglement to many promising quantum technologies such as  quantum information processing, quantum cryptography, and etc. \cite{Duan00,Ficek,WMZhang1,Guadi09,Wilk10,Rudner11,Shapiro11,Xu11,Leshem,Mundarain,Ali,Flem-Hu12}, it is of interest to investigate the sensitivity of entanglement dynamics with respect to the perturbation applied. The
relations among the entanglement dynamics, detuning between system and environment, and
the quantum memory time of the environment are studied.

The paper is organized as follows. The exact NMQSD equation is briefly reviewed in Section II, we introduce the functional expansion and the weak-coupling expansion for the NMQSD equation, and discuss the relation between them. In Section III,  a bipartite system is investigated as an example. Notably, this model allows the exact NMQSD equation. The weak-coupling approximation and the zeroth-order approximation are applied and compared with the exact solution. In addition, an analytical solution for the master equation derived for a certain group of initial states is presented.  In Section IV, based on the zeroth-order approximation, we discuss the entanglement evolution assist by a non-Markovian environment, including the generation at early stage, as well as the entanglement in the final state. The conclusion and outlook are given in Section V.

\section{NMQSD Equation and perturbation methods}
\label{sec2}
\subsection{The Non-Markovian QSD Equation}
\label{NMQSD}
A quantum open system interacting with a bosonic environment may be described by
the following total Hamiltonian,
\beqa
H_{total}&=&H_s+H_b+H_{int},H_b=\sum_i\omega_ia_i^\da a_i,H_{int}=\lambda L\sum_ig_ia_i^\da+H.c.,
\eeqa
where $H_s$ and $H_b$ are the Hamiltonian of the system of interest and the bosonic
environment, respectively. The interaction between them is described by $H_{int}$. $L$ is the
coupling operator, and $\lambda$ is the dimensionless coupling strength. Throughout this
paper, we set $\hbar=1$, for simplicity.  At zero temperature ($T=0$), the exact NMQSD equation is
given by \cite{Diosi,Yu-Strunz99, Strunz-Yu04},
\beq \label{NMQSDeq}
\frac{\partial}{\partial t}|\psi_t\rangle=[-iH_s+\lambda Lz_t^*-\lambda
L^\da\int_0^tds\alpha(t,s)\hat{O}(t,s,z^*)]|\psi_t\rangle,
\eeq
where $z^*_t=-i\sum_ig_iz_i^*e^{i\omega_it}$ is a coloured complex Gaussian process
satisfying $\mathcal{M}[z_t^*z_s^*]=0$ and $\mathcal{M}[z_tz_s^*]=\alpha(t,s)$.
$\alpha(t,s)=\sum_i|g_i|^2e^{-i\omega_i(t-s)}$ is the correlation function of the bath. Note that
$\mathcal{M}[.]$ here stands for the ensemble average over the noise $z_t^*$. The solution
of the above NMQSD equation can recover the reduced density matrix (RDM) of the system from the ensemble average
$\rho_s(t)=\mathcal{M}[|\psi_t\rangle\langle\psi_t|]$. The $O$-operator in the above equation satisfies the consistency condition $\frac{\partial}{\partial t}\frac{\delta}{\delta
z^*_s}|\psi_t\rangle=\frac{\delta}{\delta z^*_s}\frac{\partial}{\partial t}|\psi_t\rangle$, from which the equation
of motion is then derived as \cite{Yu-Strunz99}
\beq \label{consist}
\frac{\partial}{\partial t}\hat{O}(t,s,z^*)=[-iH_s+\lambda Lz_t^*-\lambda
L^\da\bar{O}(t,z^*),\hat{O}(t,s,z^*)]-\lambda L^\da\frac{\delta}{\delta z^*_s}\bar{O}(t,z^*),
\eeq
where $\bar{O}(t,z^*)=\int^t_0ds\alpha(t,s)\hat{O}(t,s,z^*)$. For the linear QSD equation above, each single realization solved from Eq.~(\ref{NMQSDeq}) does not preserve the norm, but the assemble average of many realizations converge to the RDM which is normalized. For efficient numerical simulation,  one can use the nonlinear QSD equation introduced in \cite{Diosi}:
\beqa \label{nonLNMQSDeq}
\frac{\partial}{\partial t}|\tilde{\psi}_t\rangle&=&-iH_s|\tilde{\psi}_t\rangle+\lambda (L-\langle L\rangle_t)\tilde{z}_t^*|\tilde{\psi}_t\rangle\non\\
&&-\lambda \int_0^tds\alpha(t,s)[(L^\da-\langle L^\da\rangle_t)\hat{O}(t,s,\tilde{z}^*)-\langle (L^\da-\langle L^\da\rangle_t)\hat{O}(t,s,\tilde{z}^*)\rangle]|\tilde{\psi}_t\rangle,
\eeqa
where the normalized state is defined as
\beq
|\tilde{\psi}_t\rangle=\frac{|\psi_t\rangle}{|||\psi_t\rangle||}.
\eeq
$\tilde{z}_t^*$ here is the shifted noise
\beq
\tilde{z}_t^*=z_t^*+\int_0^tds\alpha(t,s)\langle L^\da\rangle_s.
\eeq
It should be noted that all the numerical simulations presented in this paper are performed with the nonlinear version of a QSD equation [see, Eq.~(\ref{nonLNMQSDeq})].

\subsection{Functional Expansion}
\label{sec21}

In general, the $O$-operator contains the noise $z^*$ in a nonlocal way. Consider the
functional expansion of the $O$-operator in terms of $z^*_v$ \cite{Yu-Strunz99},
\beqa
\label{functional}\non
\hat{O}(t,s,z^*)&=&\hat{O}_0(t,s)+\int^t_0\hat{O}_1(t,s,v)z_v^*dv+\int^t_0\int^t_0\hat{O}_2(t,s,v_1,v_2)z_{v1}^*z_{v2}^*dv_1dv_2\\
&+&...+\int^t_0...\int^t_0\hat{O}_n(t,s,v_1,...v_n)z_{v1}^*...z_{v_n}^*dv_1...dv_n+....
\eeqa
Each term in the above expansion satisfies
\beqa\label{Consistn}\non
\frac{\partial}{\partial t}\hat{O}_n(t,s,v_1,...,v_n)&=&-i[H_s,\hat{O}_n(t,s,v_1,...,v_n)]-(n+1)\lambda L^\da\bar{O}_{n+1}(t,s,v_1,...,v_n)\\
&&-\frac{1}{n!}\sum_{P_n\in S_n}\sum^n_{k=0}[\lambda L^\da\bar{O}_{n-k}(t,v_{P_n(1)},...,v_{P_n(k)}),\hat{O}_k(t,s,v_{P_n(k+1)},...,v_{P_n(n)})],\\
&&\hat{O}_n(t,s,t,v_2,...,v_n)=\frac{1}{n}[\lambda L,\hat{O}_{n-1}(t,s,v_1,...,v_{n-1})],
\eeqa
with the initial conditions $\hat{O}_0(t,t)=\lambda L$ and $\hat{O}_n(t,t,v_1,...v_n)=0$ for
$n\geq1$. By our convention, the condition $\hat{O}(t,s,...v_m,...v_n,...)=\hat{O}(t,s,...v_n,...v_m,...)$
is valid. The necessary condition for the $O$-operator to contain a finite number of noise-dependent terms in this expansion is
\beq
[L,...[L,[L,H_s]...]]=0.
\eeq
This commutation relation can be satisfied for some interesting cases, such as dephasing model or angular momentum model with multi-level atoms \cite{Jing-Yu12}. In general, one should not expect that the functional expansion always has a finite number of noise terms as shown in the spin-boson model where an infinite number of noise terms exist \cite{You-Pert}.
A simple, yet useful approximation for numerical calculations is to truncate the $O$-operator to the zeroth order of noise
\beq\label{zero}
\hat{O}(t,s,z^*)\approx\hat{O}_0(t,s).
\eeq
For many interesting cases, it has been shown that a single realization can provide a fast estimation of the coherence information contained in the system dynamics.
In the case of approximate NMQSD, it is known that the approximate single realization would display a similar tendency as the single realization from the
exact NMQSD. More importantly, in this paper, we will systemically investigate the correction to this zeroth-order approximation by introducing higher orders of noise.

\subsection{Expansion in Terms of Coupling Strength}
\label{sec22}
Similar to the perturbation theory in quantum mechanics, the expansion of $O$-operator can be made
in terms of the coupling strength $\lambda$ \cite{Strunz01,Strunz-Yu04},
\beq\label{wcexp}
\hat{O}(t,s,z^*)=\sum_{n=1}^\infty\lambda^n\hat{O}^{(n)}(t,s,z^*).
\eeq
If the interaction between the system and its environment is weak, which is true in many
realistic cases, the weak-coupling approximation can be applied, where only the first several terms with
lower orders of $\lambda$ are considered. From Eq.~(\ref{consist}), the equation of motion for each term
in the above expansion can be derived:
\beqa\label{1}\non
\frac{\partial}{\partial t}\hat{O}^{(1)}(t,s)&=&[-iH_s,\hat{O}^{(1)}(t,s)],\\\non
\frac{\partial}{\partial
t}\hat{O}^{(2)}(t,s,z^*)&=&[-iH_s,\hat{O}^{(2)}(t,s,z^*)]+[Lz_t^*,\hat{O}^{(1)}(t,s)]-L^\da\frac{\delta}{\delta
z^*_s}\bar{O}^{(1)}(t,z^*),\\
\frac{\partial}{\partial t}\hat{O}^{(n)}(t,s,z^*)&=&[-iH_s,\hat{O}^{(n)}(t,s,z^*)]
+[Lz_t^*,\hat{O}^{(n-1)}(t,s,z^*)]\\\non
&&-\sum_{k=1}^{n-2}[L^\da\bar{O}^{(k)}(t,s,z^*),\hat{O}^{(n-1-k)}(t,s,z^*)]-L^\da\frac{\delta}{\delta
z^*_s}\bar{O}^{(n-1)}(t,z^*),
\eeqa
with the boundary conditions $\hat{O}^{(1)}(t,t)=L$ and $\hat{O}^{(n)}(t,t,z^*)=0$ for
$n>1$. Note that the solution of the first term in the above expansion is given by
\beq
\hat{O}^{(1)}(t,s)=e^{-iH_s(t-s)}Le^{iH_s(t-s)},
\eeq
which does not contain noise $z_t^*$. Furthermore, if the commutation relation $[L, \hat{O}^{(n)}(t,s,z^*)]=0$ is valid, then the next term $\hat{O}^{(n+1)}(t,s)$ does not contain noise. However, in general such condition is not true. Notably, the order
of noise is only dependent on the commutator  $[L, \hat{O}^{(n)}(t,s,z^*)]$, thus for each
$\hat{O}^{(n)}$ associated with $\lambda^n$, the order of noise contained in this term is always lower than $n$.

\subsection{Functional Expansion Versus Weak-Coupling Expansion}
\label{FvW}

Given the two perturbation methods above, it is interesting to investigate the relation between the two expansions. The zeroth-order approximation can provide many useful results for non-Markovian systems and at the same time greatly reduce the complexity of the numerical simulations. But up to now,  the physical meaning behind the functional expansion has not been discussed in details. It is known that the weak-coupling expansion has a clear physical interpretation in terms of coupling strength.
Therefore, by comparing the two expansions, we systematically show the broad range of availability for the zeroth-order approximation. To do so, we further expand each term in the functional expansion in terms of the coupling strength $\lambda$,
\beqa
\label{weakcoupling}
\hat{O}_n(t,s,v_1,...v_n)=\sum_{m=1}^\infty\lambda^m\hat{O}_n^{(m)}(t,s,v_1,...v_n).
\eeqa
For example, in the zeroth-order approximation Eq.~(\ref{zero}), $\hat{O}_0(t,s)$ can be further expanded  as
\beq
\hat{O}_0(t,s)=\lambda\hat{O}_0^{(1)}(t,s)+\lambda^2\hat{O}_0^{(2)}(t,s)+\lambda^3\hat{O}_0^{(3)}(t,s)....
\eeq
Substituting this expansion into the equation of motion Eq.~(\ref{Consistn}), we obtain
\beq
\frac{\partial}{\partial t}\hat{O}_0^{(n)}(t,s)=-i[H_s,\hat{O}_0^{(n)}(t,s)]
-\sum_{k=1}^{n-1}[L^\da\hat{O}_0^{(k)}(t,s),\hat{O}_0^{(n-1-k)}(t,s)].
\eeq
With the initial condition $\hat{O}_0(t,t)=\lambda L$, one can show $\hat{O}_0^{(n)}(t,t)=0$ if $n$ is even. Therefore,
\beq
\hat{O}_0(t,s)=\lambda\hat{O}_0^{(1)}(t,s)+\lambda^3\hat{O}_0^{(3)}(t,s)+\lambda^5\hat{O}_0^{(5)}(t,s)+\cdots.
\eeq
Note that the first term $\hat{O}_0^{(1)}(t,s)$ above is the same as the first term in the
weak-coupling expansion, i.e. $\hat{O}_0^{(1)}(t,s)=\hat{O}^{(1)}(t,s)$. Therefore, the approximation $\hat{O}(t,s,z^*)\approx\hat{O}_0(t,s)$ typically contains more dynamical information  than $\hat{O}^{(1)}(t,s)$, since the the later term is already included in the zeroth-order expansion. However, as discussed in Sec.~\ref{sec22}, in general, $\hat{O}^{(2)}(t,s,z^*)$ may contain linear noise, thus it is not conclusive that $\hat{O}_0(t,s)$ defined in the functional expansion is always better than the second-order weak-coupling expansion. Nonetheless, for certain models where  the condition $[L,\hat{O}^{(1)}(t,s)]=0$ holds, then $\hat{O}^{(2)}(t,s,z^*)=0$. For those models, $\hat{O}_0(t,s)$ in the functional expansion will be more accurate than the second-order weak-coupling expansion of the $O$-operator, as shown in the example below.

We consider a system consisting of an atom with $m$ levels, where the Hamiltonian can be written in the following form,
\beqa
H_s=\sum_{i=1}^{m}C_iH_s^{(i)},\ L=\sum_{j=2}^{m}G_jL^{(j)}.
\eeqa
Here $H_s^{(i)}=|i\rangle\langle i|$, $L^{(j)}=|j-1\rangle\langle j|$. For a spin-$l$
system, we have $m=2l+1$ and $C_i=(-l-1+i)\omega$, where $\omega$ is the energy difference between
two neighboring levels. The first term $\hat{O}^{(1)}$ in the expansion can be solved with
\beq\non
[H_s, L]=\sum_{i=2}^{m}(C_{i-1}-C_{i})G_i|i-1\rangle\langle i|=-\omega L.
\eeq
Then $\hat{O}^{(1)}(t,s)=f(t,s)L$, where $f(t,s)$ can be solved from Eq.~(\ref{1}). Obviously, here $[\hat{O}^{(1)}(t,s), L]=0$. Based on Eq.~(\ref{1}), the second term in the expansion satisfies
\beq
\frac{\partial}{\partial t}\hat{O}^{(2)}(t,s,z^*)=[-iH_s,\hat{O}^{(2)}(t,s,z^*)],
\eeq
and the solution is thus $\hat{O}^{(2)}(t,s,z^*)=0$ due to the boundary condition. Subsequently, $\hat{O}^{(3)}(t,s)$ must be noise-free. Therefore, all the remaining noise terms must be proportional to a term with $\lambda^4$ or
higher order terms (as shown in Eq.~(\ref{wcexp})). This conclusion can be easily generalized
to the case of multiple atoms. With $[H_{sA}^{(i)}, L^{(j)}_B]=0$ for different atoms $A$ and
$B$, the conclusion for the approximation still holds.

The relation between the weak-coupling and the functional expansion is summarized in Table.~\ref{table1}.  Each column in the table represents a term in the functional expansion Eq.~(\ref{functional}), which is further expanded in terms of coupling strength $\lambda$.  Each row represents a term in the coupling strength expansion Eq.~(\ref{weakcoupling}).
Note that each term $\hat{O}_n^{(m)}(t,s,z^*)$ in the expansion may contains noise, but the order of the noise associated with it can never go higher than the order of $\lambda$, thus the entries in up-right corner of the table are all zeros. The exact $O$-operator contains  all the non-zero terms in this table. Although it is still not clear about the convergence property of the two expansions, our example shows that the functional expansion typically includes more terms than the same order coupling-strength expansion. Namely, for the given order the functional expansion should gives rise to a better approximation than the coupling-strength expansion. Some details about this comparison will be given below in our discussions on a dissipative model.

\begin{table}[htbp]\centering
\begin{tabular}{|c|c|c|c|c|} \hline
$\hat{O}(t,s,z_t^*)$ & $\hat{O}_0(t,s)$ & $\hat{O}_1(t,s,v)$ & $\hat{O}_2(t,s,v_1,v_2)$ & ...\\ \hline
$\lambda\hat{O}^{(1)}(t,s)$ & $\hat{O}_0^{(1)}$ & 0  & 0 & ...\\ \hline
$\lambda^2\hat{O}^{(2)}(t,s,z_t^*)$ & $ 0 $ & $\hat{O}_1^{(2)}$ & 0 & ...\\ \hline
$\lambda^3\hat{O}^{(3)}(t,s,z_t^*)$ & $\hat{O}_0^{(3)}$ & $\hat{O}_1^{(3)}$ & $\hat{O}_2^{(3)}$ & ...\\ \hline
... & ...& ...& ...& ...\\ \hline
\end{tabular}
\caption{Relation between the functional expansion and the weak-coupling expansion.}
\label{table1}\end{table}

\section{Two-Qubit Dissipative Model}
\label{sec3}

\subsection{Exact Solution for a Bipartite System}
\label{sec31}
Based on the equation of motion for the $O$-operator given in the previous section, here we present the NMQSD equation for a specific example. The model considered here is two uncoupled qubits interacting with a common dissipative bosonic environment. For this example, the exact NMQSD equation is available and the $O$-operator contains only the first-order noise, so that we can compare each approximation with the exact NMQSD equation. The Hamiltonian for this dissipative model is given by
\beqa
H_s&=&\frac{\omega_A}{2}\si_z^A+\frac{\omega_B}{2}\si_z^B,\\
H_{int}&=&\lambda\sum_i(g_iLa_i^\da+g_i^*L^\da a_i),
\eeqa
where $L=\si_-^A+\si_-^B$. For simplicity we consider $\omega_A=\omega_B=\omega_s$, and the model with different frequencies
can be solved similarly. The
exact $O$-operator in the NMQSD equation can be found using the functional expansion \cite{Xinyu},
\beq\label{exacto}
\bar{O}(t,z^*)=\bar{O}_0(t)+\int^t_0\bar{O}_1(t,v)z_v^*dv,
\eeq
where $\bar{O}_0(t)=F_1(t)(\si_-^A+\si_-^B)+F_2(t)(\si_z^A\si_-^B+\si_-^A\si_z^B)$,
$\bar{O}_1=F_3(t,v)\si_-^A\si_-^B$. We choose the spectrum of the environment to be
Lorentz-type
\beq
J(\omega)=\frac{\Gamma}{\pi}\frac{\gamma^2}{(\omega-\Omega)^2+\gamma^2},
\eeq
where $\gamma^{-1}$ is the environmental memory time, $\Omega$ is the central frequency of the
environment, and we assume $\Gamma=1$. Since the noise is a complex
Ornstein-Ulenbeck type, then the correlation function can be given by
\beqa
\alpha(t,s)=\int_0^\infty d\omega J(\omega)e^{-i\omega(t-s)}=\frac{\gamma}{2}e^{-\gamma|t-s|}e^{-i\Omega(t-s)}.
\eeqa
From the consistency
condition, the equations for coefficients $F_1$, $F_2$, and $F_3$ can be derived,
\beqa\non
\dot{F}_1(t)&=&\frac{\lambda\gamma}{2}-\gamma F_1(t)+i\Delta F_1(t)+\lambda F_1(t)^2+3\lambda
F_2(t)^2-i\frac{1}{2}\lambda\bar{F}_3(t),\\
\dot{F}_2(t)&=&-\gamma F_2(t)+i\Delta F_2(t)-\lambda F_1(t)^2+4\lambda F_1(t)F_2(t)+\lambda
F_2(t)^2-i\frac{1}{2}\lambda\bar{F}_3(t),\\
\dot{\bar{F}}_3(t)&=&-2\gamma \bar{F}_3(t)+i2\Delta\bar{F}_3(t)+4\lambda F_1(t)\bar{F}_3(t)-i2\gamma\lambda
F_2(t),\\
F_3(t,v)&=&-4iF_2(v)\exp[\int^t_{v}ds (-\gamma+i2\omega_s-i\Omega+4\lambda F_1(s))],
\label{F123}
\eeqa
where $\Delta=\omega_s-\Omega$ is the detuning between the system and the environment. The
dynamics of the system can be obtained by solving the nonlinear NMQSD equation Eq.~(\ref{nonLNMQSDeq}) with the coefficients above. In what follows, we apply both perturbation methods to this model, and compare the results with the exact solution. Clearly, the rigorous quantification of the accuracy of the approximations is a very complex  task. We use Wootter's concurrence \cite{Wootters} and fidelity as the indicator to show the deviation of the approximated results from the exact ones. As shown below, the approximations give a better result when the noise coupling is weak.

\subsection{Weak-Coupling Approximation}

\begin{figure}[!t]
\includegraphics[width = 10cm]{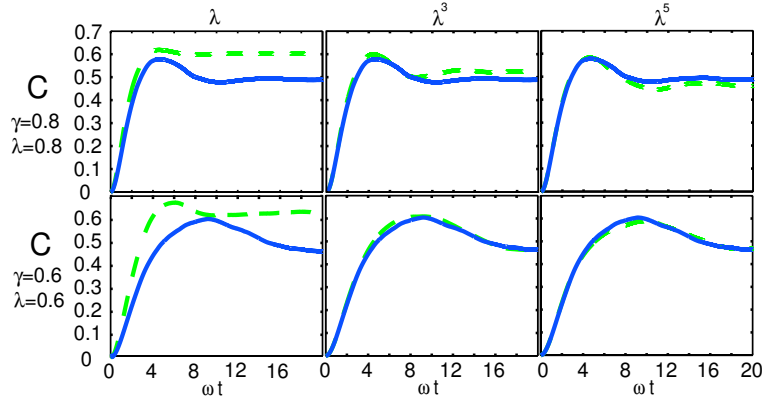}
\caption{(Color online) Concurrence evolution of the bipartite system with
the initial state $|01\rangle$. The $O$-operator is expanded in terms of coupling strength up
to $\lambda$, $\lambda^3$, $\lambda^5$. Blue solid lines are plotted from the exact solution, green
dashed lines are plotted from the weak-coupling approximation.}\label{wcC}
\end{figure}

If the coupling strength $\lambda$ is small, the NMQSD equation obtained from the
weak-coupling approximation can well describe the dynamics of the quantum open system \cite{Strunz01,Strunz-Yu04}. From
Sec.~\ref{sec22}, the first five terms of the expansion Eq.~(\ref{wcexp}) are given by
\beqa\non
\hat{O}^{(1)}(t,s)&=&\si_-^Ae^{i\omega_s(t-s)}+\si_-^Be^{i\omega_s(t-s)},\\\non
\hat{O}^{(2)}(t,s)&=&0,\\\non
\hat{O}^{(3)}(t,s)&=&f_3^A(t,s)\si_-^A+f_3^B(t,s)\si_-^B+g_3^A(t,s)\si_z^A\si_-^B+g_3^B(t,s)\si_-^A\si_z^B,\\\non
\hat{O}^{(4)}(t,s,z^*)&=&h_4(t,s,z_t^*)\si_-^A\si_-^B,\\\non
\hat{O}^{(5)}(t,s)&=&f_5^A(t,s)\si_-^A+f_5^B(t,s)\si_-^B+g_5^A(t,s)\si_z^A\si_-^B+g_5^B(t,s)\si_-^A\si_z^B,
\eeqa
where the time dependent coefficients above can be solved from Eq.~(\ref{1}). The details of the derivation are shown in Appendix \ref{A1}. We can show for $n=1,2,...,\infty$,
\beqa
\hat{O}^{2n}(t,s,z^*)&=&\int_0^tds'h_{2n}(t,s,s')z_{s'}\si_-^A\si_-^B,\\
\hat{O}^{2n-1}(t,s)&=&f_{2n-1}^A(t,s)\si_-^A+f_{2n-1}^B(t,s)\si_-^B+g_{2n-1}^A(t,s)\si_z^A\si_-^B+g_{2n-1}^B(t,s)\si_-^A\si_z^B,
\eeqa
which indicate that only the first-order noise exists in $\hat{O}(t,s,z^*)$ and they only associate with even orders of $\lambda$, as discussed in Sec.~\ref{FvW}. This result is consistent with the exact solution. Fig.~\ref{wcC} plots the concurrence evolution of the bipartite system with different orders of the coupling strength truncations (dashed lines), and compared to the exact solution (solid lines). The initial state of the system is $|01\rangle$, which will result in an entangled final state. The figures on the upper panel show the case when the coupling is strong ($\gamma=0.8\omega_s$), and the approximate result
is closer to the exact solution if we include more terms in the expansion of $O$-operator. Notably, the residue entanglement calculated
from the weak-coupling approximation is different from the exact solution. In the figures on the lower panel, the coupling strength $\lambda$ is reduced to $0.6$, which can still be considered as a strong coupling regime, and the result is already improved. Here $\gamma$ is chosen to be $0.6\omega_s$, which introduces more pronounced non-Markovian feature to the dynamics. The validity of the weak-coupling approximation is also related to the quantum memory time $\gamma^{-1}$ of the bath, but less sensitive comparing to $\lambda$. In the Markov limit where $\gamma\rightarrow\infty$, we have $\bar{O}(t)=\lambda\bar{O}^{(1)}(t)=\lambda L$.

\subsection{Zeroth-Order Approximation}
\label{zona}

\begin{figure}[!t]
\includegraphics[width = 8cm]{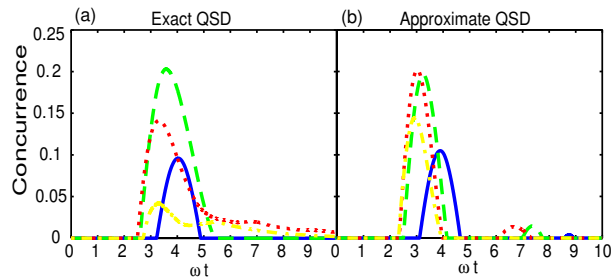}
\caption{(Color online) Generation of entanglement from $|11\rangle$ state with different $\gamma$. $\omega_s=1$, $\gamma=0.2$ (Blue Solid), $0.5$ (Green dashed), $0.8$ (Red dotted), and $1.1$ (Yellow dashed dotted). $\Delta=1$. $(a)$ is plotted from the exact NMQSD, $(b)$ is plotted from the zeroth-order approximation.}\label{11}
\end{figure}

The second approximation to be investigated is the zeroth-order approximation based on the functional expansion Eq.~(\ref{zero}) \cite{Yu-Strunz99}. Based on the the exact $O$-operator given by Eq.~(\ref{exacto}), when the zeroth-order approximation is applied, the noise term to be truncated is $\bar{O}_1(t,v)=F_3(t,v)\si_-^A\si_-^B$.  It is easy to check that
the operator $\bar{O}_1(t,v)$ annihilates the following state vectors $|01\rangle, |10\rangle, |00\rangle$, and the population in the $|11\rangle$ state can never increase (See Appendix~\ref{A2}). Therefore, the noise term in the $O$-operator will not affect the initial states that do not contain $|11\rangle$.
Since the main purpose here is to discuss how an approximation affects non-Makovian dynamics, we consider the worst scenario where the
initial state is $|11\rangle$, which should result in the most significant deviation from the exact solution comparing to other initial states. The difference is examined by comparing the entanglement dynamics calculated from the approximated NMQSD with that from the exact NMQSD equation Eq.~(\ref{NMQSDeq}). The results are shown in Fig.~\ref{11}, where the concurrence from the exact NMQSD equation is plot in Fig.~\ref{11} $(a)$, and that from the approximated NMQSD equation is in $(b)$. Once can notice the slight differences between the two figures, since the noise term $\hat{O}_1(t,s,z^*)$ in the $O$-operator contains further dynamical information. Note that the generated concurrence is not monotonically related to the memory time \cite{Xinyu}. This interesting non-Markovian feature is also captured by the approximate NMQSD equation. This example has shown that, for the given parameters, the zeroth-order expansion is a very good approximation for the dissipative model in non-Markov regime.

\begin{figure}[!t]
\includegraphics[width = 8cm]{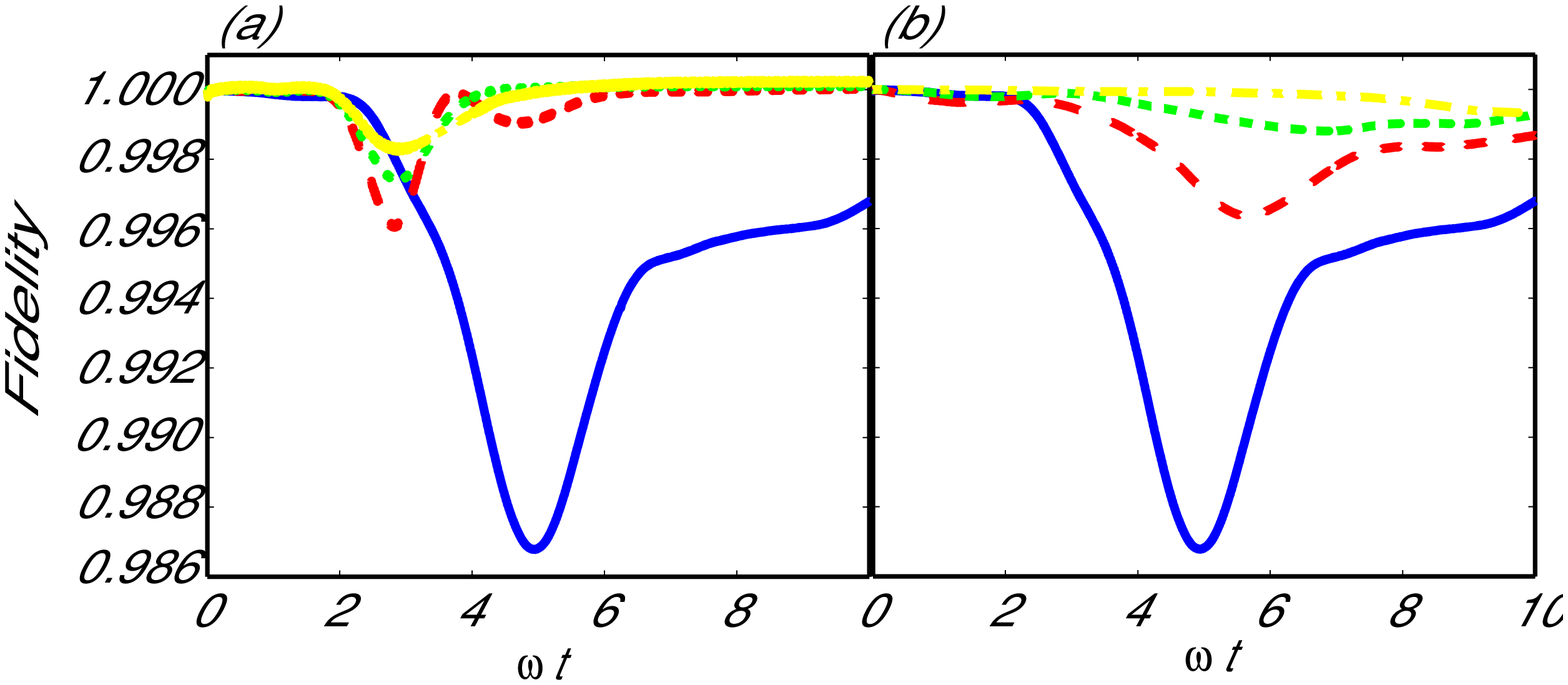}
\caption{(Color online) Fidelity between the $\rho_s(t)$ from the exact NMQSD and the approximated master equation. The initial state is $|11\rangle$ state, $\omega_s=1$, $\Delta=1$. $(a)$ Fixed $\lambda=1$,
$\gamma=0.5$ (Blue Solid), $1.0$ (Red dashed), $1.5$ (Green dotted), $1.1$ (Yellow dashed
dotted). $(b)$ Fixed  $\gamma=0.5$, $\lambda=1.0$ (Blue Solid), $0.8$ (Red dashed), $0.6$ (Green dotted), $0.4$ (Yellow dashed
dotted).}\label{F}
\end{figure}

To further investigate the zeroth-order approximation, we derive the corresponding non-Markovian master equation
\beq \label{ME}
\dot{\rho}_s(t)=-i[H_s,\rho_s(t)]+\lambda[L,\rho_s(t)\bar{O}_0^\da(t)]+\lambda[\bar{O}_0(t)\rho_s(t),L^\da].
\eeq
The accuracy of this zeroth-order master equation can be examined with the fidelity defined below,
\begin{equation}
F(t)=\text{Tr}\sqrt{\sqrt{\rho_{\text{ex}}(t)}\rho_{\text{app}}(t)\sqrt{\rho_{\text{ex}}(t)}},
\end{equation}
where $\rho_{\text{ex}}(t)$ and $\rho_{\text{app}}(t)$ are obtained from the exact NMQSD Eq.~(\ref{NMQSDeq}) and the approximated master equation Eq.~(\ref{ME}), respectively. The initial state considered here is again $|11\rangle$, which is known to be sensitive to the approximation. The result is shown in Fig.~\ref{F}. As expected, the accuracy is shown to be proportional to $\gamma$. That is, the approximation becomes better in a stronger Markov regime or the coupling strength is weaker. However, even in the non-Markovian regime with a strong coupling constant, the approximation is still very close to the exact solution. The lowest fidelity in the figure is around $0.986$, showing that the approximation is still reliable in that parameter range.

Next, we consider a set of special initial states where the population in $|11\rangle$ is zero. Since the population in $|11\rangle$ state decays exponentially due to the dissipative interaction, any initial state eventually evolves into this set of states. Under such initial condition, the matrix form of the RDM remain as
\beqa
\rho_s(t)=\left(\begin{array}{cccc} 0 & 0 & 0 & 0\\
0 & \rho_{22}(t) & \rho_{23}(t) & \rho_{24}(t) \\0 & \rho_{32}(t) & \rho_{33}(t) &
\rho_{34}(t) \\
0 & \rho_{42}(t) & \rho_{43}(t) & \rho_{44}(t) \end{array} \right).
\eeqa
An analytical solution of the above RDM can be derived, as shown in
Appendix~\ref{A2}. The concurrence for the RDM in such a form is given by(t)
\beqa \label{analy}
C(t)&=&2|\rho_{23}(t)|=2\sqrt{R_{23}^2(t)+I_{23}^2(t)},\\
R_{23}(t)&=&\frac{1}{4}(|A(t)|^2-1)\rho_{22}(0)+\frac{1}{4}(|A(t)|^2-1)\rho_{33}(0)+\frac{1}{2}(|A(t)|^2+1)R_{23}(0),\\
I_{23}(t)&=&\frac{1}{2}\Im[A(t)](\rho_{33}(0)-\rho_{22}(0))+\Re[A(t)]I_{23}(0),\\
A(t)&=&e^{-2\int_0^t X(s) ds}=\frac{2\sqrt{\gamma}}{\beta}e^{-\frac{1}{2}\gamma t}e^{i\frac{1}{2}\Delta
t}\cos(\frac{1}{2}\beta t+\arctan\frac{i\Delta-\gamma}{\beta}),\\
X(t)&=&\frac{1}{4\lambda}[\gamma-i\Delta+\beta\tan(\frac{1}{2}\beta t+c)],
\eeqa
where $X(t)=F_1(t)-F_2(t)$, $\beta=\sqrt{4\lambda^2\gamma-\gamma^2+i2\gamma\Delta+\Delta^2}$,
$c=\tan^{-1}\left(\frac{i\Delta-\gamma}{\beta}\right)$. $R_{23}(t), I_{23}(t)$ are the real and imaginary part of $\rho_{23}(t)$ respectively. From the analytical solution, the dynamics of the matrix elements in the RDM can be completely described by the  function $A(t)$. The exponential decay rate is proportional to $\frac{\gamma}{2}$, while the oscillation is
determined by the detuning $\Delta$ and $\gamma$ together. These oscillation terms are
responsible for the non-monotonic dynamics of the RDM at early times.

\begin{figure}[!t]
\includegraphics[width = 8cm]{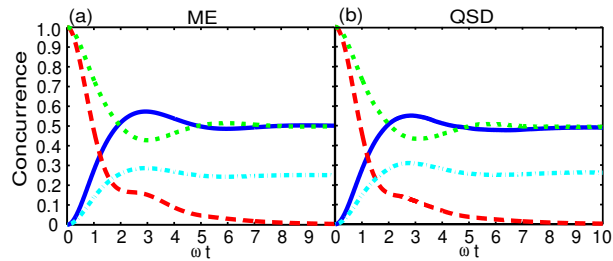}
\caption{The concurrence evolution for initial state $|10\rangle$ (blue solid line),
$\frac{1}{\sqrt2}(|10\rangle+|01\rangle)$ (red dashed line),
$\frac{1}{\sqrt2}(|10\rangle-i|01\rangle)$ (green dotted line), and
$\frac{1}{\sqrt2}(|10\rangle+|00\rangle)$ (cyan dash-dotted line). $(a)$ is plotted from the zeroth-order master equation, $(b)$ is plotted from the exact NMQSD.}\label{Cana}
\end{figure}

Fig.~\ref{Cana} compares the results from the zeroth-order master equation Eq.~(\ref{ME}) and the exact NMQSD equation. All the initial states chosen here have zero population in $|11\rangle$ state. It is shown that the results are consistent with each other perfectly. The accuracy of the approximation under these initial states is independent of the environmental parameters $\gamma$ and $\Omega$.

\section{Entanglement Assisted by a Non-Markovian Bath}
\label{entanglementgeneration}
\subsection{Entanglement Generation at Early Stage}
\label{earlystage}
\begin{figure}[!t]
\includegraphics[width = 8cm]{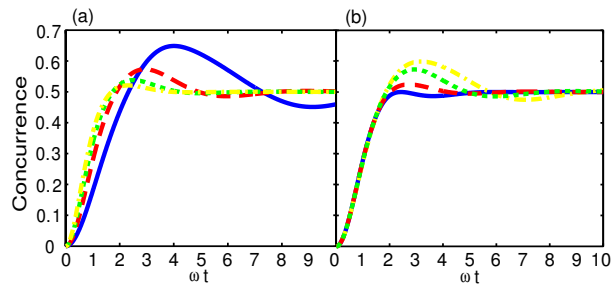}
\caption{The concurrence evolution for initial state $|10\rangle$ with different
environmental parameters. $(a)$ $\Delta=1$, $\gamma=0.5$ (blue solid line), $1.0$ (red
dashed line), $1.5$ (green dotted line), $2.0$ (yellow dash dotted line). $(b)$ $\gamma=1$,
$\Delta=0$ (blue solid lines), $0.5$ (red dashed line), $1.0$ (green dotted line), $1.5$
(yellow dash dotted line).} \label{Cmax}
\end{figure}

At an early stage, the entanglement dynamics is greatly affected by the environment memory time. The back reaction from the environment can significantly enhance the generation of entanglement, especially when the memory time $\gamma^{-1}$ is long.
Fig.~\ref{Cmax} $(a)$ shows the generation of entanglement from the separable initial state
$|10\rangle$, where the final entangled state has a concurrence $C=0.5$. The maximum instantaneous concurrence generated at early stage can benefit from the strong non-Markovian environment. Interestingly, due to the delayed
response from the environment, the times needed for the maximal entanglement to be generated is proportional to the environmental
memory time $\gamma^{-1}$. The time for reaching the maximal entanglement is longer with a longer memory time for
the environment. Moreover, the entanglement generated needs longer time to settle down to the steady state.

The detuning $\Delta$ also affects the dynamics of the system
significantly as shown in Fig.~\ref{Cmax} $(b)$. For the dissipative model, the environment
interact with the system by exchanging energy. Therefore in the resonant case, the system
will quickly approach its steady state with small fluctuations. As detuning increases,
the influence from the environment becomes weaker. Thus a stronger generation of
entanglement and other non-Markovian behaviors may be observed.

\subsection{Entanglement Residue in the Long-Time Limit}
\label{finalform}

In the long-time limit, the steady state of the system  may become entangled depending
 on the initial states of the system \cite{Mundarain,Xinyu}. Consider the long-time limit $t\gg\tau_S$,
 where $\tau_S$ is the relaxation time of the system, from the dynamic equations in Appendix \ref{A2}, the RDM will have the
following form:
\beqa\label{final}
\rho_s(t\gg\tau_S) &\approx& \left(\begin{array}{cccc} 0 & 0 & 0 & 0\\
0 & r & -r & x \\0 & -r & r & -x \\
0 & x^* & -x^* & 1-2r \end{array} \right),
\eeqa
where $x=\frac{1}{2}(\rho_{24}(0)-\rho_{34}(0))e^{-i\omega_s t}$. The value of $r$ is given by
\beq\label{condition}
r=\frac{1}{4}[\rho_{22}(0)+\rho_{33}(0)-2\Re{[\rho_{23}(0)]}].
\eeq
Thus, we find a direct relationship between the initial state and the concurrence of  the final state,
\beqa
C(t\gg\tau_S)=2r.
\eeqa
This result provides a very useful way of identifying a set of  ``good''  initial separable states that can produce desired entangled steady states. Obviously, if the final entangled states are of interest, then the initial entangled states are not always good choices. For example, the steady state of the Bell state $|\psi\rangle =\frac{1}{\sqrt{2}}(|11\rangle \pm |00\rangle)$ will have zero concurrence.

\section{Conclusion and Outlook}
\label{Conclusion}

In this paper we have discussed two perturbative expansions for the NMQSD equation based on a dissipative quantum open system.  A
detailed comparison between the functional expansion and the weak-coupling expansion is
presented by discussing the temporal evolution of a bipartite dissipative system. We systematically
study the validity ranges of the two approximations of the $O$-operator, where the determination of $O$-operator is essential for efficiently simulating the non-Markovian quantum open systems. For the specific model involving a dissipative environment considered in this paper, it is shown that the functional expansion has advantages in solving the system dynamics compared with the weak-coupling approximation, irrespective of the coupling strength. In more general cases, given the same number of terms kept in the expansion, the functional expansion should be as good as the weak-coupling expansion. However, a cut-off on the noise terms in the $O$-operator expansion may fail to capture some non-Markovian features such as faster coherence decay or lack of the entanglement generation. Generally, the higher order noise terms correspond to higher order weak coupling approximation in terms of the coupling strength $\lambda$. Particularly, for the dissipative model considered in this paper,  the noise terms contain coupling strength up to order $\lambda^4$.  Based on this observation, it is easy to see that why the zeroth-order approximation for the dissipative model can give a very accurate description. In addition, we have derived a non-Markovian master equation from an approximated NMQSD equation, from which an analytical solution is obtained. As an application of our perturbation approaches, we have discussed how the entanglement generation is related to environmental memory, coupling strength, and detuning. Clearly, it is desirable to study the relationship between the functional expansion and weak-coupling expansion for a more general system.
As a future project, we will study the higher-order perturbation and non-Markovian corrections for more generic systems.

\section*{Acknowledgement}
We acknowledge grant support from the AFOSR No. FA9550-12-1-0001.

\appendix
\section{Weak-Coupling Expansion for the Bipartite System}\label{A1}
For a bipartite dissipative system, consider $H_s=\frac{\omega_s}{2}\si_z^A+\frac{\omega_s}{2}\si_z^B$, $L=\si_-^A+\si_-^B$. Applying the equation of motion of the $O$-operaotr Eq.~(\ref{Consistn}) and the weak-coupling expansion Eq.~(\ref{wcexp}), the first three terms in the expansion are given by
\beqa\non
\hat{O}^{(1)}(t,s)&=&\si_-^Ae^{i\omega_s(t-s)}+\si_-^Be^{i\omega_s(t-s)},\\\non
\hat{O}^{(2)}(t,s)&=&0,\\
\frac{\partial}{\partial t}\hat{O}^{(3)}(t,s)&=&[-iH_s,\hat{O}^{(3)}(t,s)] -z_t^*[L,\hat{O}^{(2)}(t,s)]-[L^\da\bar{O}^{(1)}(t,s),\hat{O}^{(1)}(t,s)]-L^\da\frac{\delta}{\delta z^*_s}\bar{O}^{(2)}(t).
\eeqa
Note $\hat{O}^{(3)}(t,s)$ must be noise-free since $\hat{O}^{(2)}(t,s)=0$. Thus assume
$\hat{O}^{(3)}(t,s)=f_3^A(t,s)\si_-^A+f_3^B(t,s)\si_-^B+g_3^A(t,s)\si_z^A\si_-^B+g_3^B(t,s)\si_-^A\si_z^B$, then
\beqa\non
\dot{f}_3^A(t,s)&=&i\omega_sf_3^A(t,s)+F_1^A(t)f_1^A(t,s),\\\non
\dot{f}_3^B(t,s)&=&i\omega_sf_3^B(t,s)+F_1^B(t)f_1^B(t,s),\\\non
\dot{g}_3^A(t,s)&=&i\omega_sg_3^A(t,s)-f_1^A(t,s)F_1^B(t),\\\non
\dot{g}_3^B(t,s)&=&i\omega_sg_3^B(t,s)-f_1^B(t,s)F_1^A(t),
\eeqa
with initial condition $F_n^j(0)=0,\ G_n^j(0)=0$ and $F_n^j(t)=\int_0^tds\alpha(t,s)f_n^j(t,s)$, $j=A,B, n=1,3$. Then $\hat{O}^{(4)}(t,s,z_t^*)=h_4(t,s,z_t^*)\si_-^A\si_-^B$,
\beqa\non
\frac{\partial}{\partial t}\hat{O}^{(4)}(t,s,z_t^*)&=&[-iH_s,\hat{O}^{(4)}(t,s,z_t^*)]+[Lz_t^*,\hat{O}^{(3)}(t,s)]\\\non
&&-[L^\da\bar{O}^{(1)}(t,s),\hat{O}^{(2)}(t,s)]-[L^\da\bar{O}^{(2)}(t,s),\hat{O}^{(1)}(t,s)]-L^\da\frac{\delta}{\delta z^*_s}\bar{O}^{(3)}(t)\\
&=&[-iH_s,\hat{O}^{(4)}(t,s,z_t^*)]+[Lz_t^*,\hat{O}^{(3)}(t,s)].
\eeqa
We can solve this by assuming $h_4(t,s,z_t^*)=\int_s^t ds'h'_4(t,s,s')z_{s'}^*$, and with $H'_4(t,s')=\int_0^tds\alpha(t,s)h'_4(t,s,s')$. Pluging them into the left and right side of the equation above, one can obtain
\beqa\non
&&\frac{\partial}{\partial t}\int ds'H'_4(t,s')z_{s'}^*=i2\omega_s\int_s^t ds'H'_4(t,s')z_{s'}^*+2z_t^*(G_3^A(t)+G_3^B(t)),\\\non
&&H'_4(t,t)=2(G_3^A(t)+G_3^B(t)),\\\non
&&H'_4(t,s')=2(G_3^A(s')+G_3^B(s'))e^{(-R+i2\omega_s))(t-s')}.
\eeqa
Furthermore, since $[L,\hat{O}^{(4)}(t,s,z_t^*)]=0$, from
\beqa
\frac{\partial}{\partial t}\hat{O}^{(5)}(t,s)&=&[-iH_s,\hat{O}^{(5)}(t,s)]-[L^\da\bar{O}^{(1)}(t,s),\hat{O}^{(3)}(t,s)]
-[L^\da\bar{O}^{(3)}(t),\hat{O}^{(1)}(t,s)]-L^\da\frac{\delta}{\delta
z^*_s}\bar{O}^{(4)}(t,z_t^*),
\eeqa
$\hat{O}_5(t,s)$ must be noise free. Then one can obtain $\hat{O}_5(t,s)=f_5^A(t,s)\si_-^A+f_5^B(t,s)\si_-^B+g_5^A(t,s)\si_z^A\si_-^B+g_5^B(t,s)\si_z^B\si_-^A$, with
\beqa\non
\dot{f}_5^A(t,s)&=&i\omega_sf_5^A(t,s)-F_1^A(t)(g_3^A(t,s)-f_3^A(t,s))+f_1^A(t,s)(F_3^A(t)+G_3^A(t))-\frac{1}{2}H'_4(t,s),\\\non
\dot{f}_5^B(t,s)&=&i\omega_sf_5^B(t,s)-F_1^B(t)(g_3^B(t,s)-f_3^B(t,s))+f_1^B(t,s)(F_3^B(t)+G_3^B(t))-\frac{1}{2}H'_4(t,s),\\\non
\dot{g}_5^A(t,s)&=&i\omega_sg_5^A(t,s)-F_1^B(t)(f_3^A(t,s)-g_3^A(t,s))-\frac{1}{2}H'_4(t,s)\\\non
&&-(F_3^B(t)f_1^A(t,s)-G_3^A(t)f_1^A(t,s)-G_3^A(t)f_1^B(t,s)-G_3^B(t)f_1^B(t,s)),\\\non
\dot{g}_5^B(t,s)&=&i\omega_sg_5^B(t,s)-F_1^A(t)(f_3^B(t,s)-g_3^B(t,s))-\frac{1}{2}H'_4(t,s)\\\non
&&-(F_3^A(t)f_1^B(t,s)-G_3^B(t)f_1^B(t,s)-G_3^B(t)f_1^A(t,s)-G_3^A(t)f_1^A(t,s)).
\eeqa

\section{Dynamical Equations for the Matrix Elements in the RDM}\label{A2}

Based on the zeroth-order approximation Eq.~(\ref{zero}) and the corresponding master equation Eq.~(\ref{ME}), we derive the dynamical equation for each independent element in the RDM (only nine of the sixteen elements are
independent). Using the notation $X(t)=F_1(t)-F_2(t)$, $Y(t)=F_1(t)+F_2(t)$, from Eq.~(\ref{F123}) under the approximation one can obtain,
\beqa\label{XY}
\dot{X}(t)&=&(i\Delta-\gamma)X(t)+2\lambda X(t)^2+\frac{\lambda\gamma}{2},\\
\dot{Y}(t)&=&(i\Delta-\gamma-2\lambda X(t))Y(t)+2\lambda Y(t)^2+\frac{\gamma}{2},\label{Y}\\
X(t)&=&\frac{1}{4\lambda}[\gamma-i\Delta+\beta\tan(\frac{1}{2}\beta t+c)],
\eeqa
where $\beta=\sqrt{4\lambda^2\gamma-\gamma^2+i2\gamma\Delta+\Delta^2}$,
$c=\tan^{-1}\left(\frac{i\Delta-\gamma}{\beta}\right)$.
It is hard to find a solution to Eq.~(\ref{Y}) analytically. However, in the long time limit where
$X(t\rightarrow\infty)=\frac{1}{4\lambda}(\gamma-i\Delta+i\beta)$, the dynamics of $Y(t)$ will be the same as
$X(t)$. Notably, the solution for $X(t)$ is valid without the approximation. The dynamical equations for each element in the RDM is then given by
\beqa\label{dy1}
\dot{\rho}_{11}(t)&=&-4\lambda\Re[Y(t)]\rho_{11}(t),\\ \non
\dot{\rho}_{12}(t)&=&-i\omega\rho_{12}(t)-\lambda(2Y(t)+X^*(t))\rho_{12}(t)-\lambda X(t)\rho_{13}(t),\\  \non
\dot{\rho}_{13}(t)&=&-i\omega\rho_{13}(t)-\lambda(2Y(t)+X^*(t))\rho_{13}(t)-\lambda X(t)\rho_{12}(t),\\ \non
\dot{\rho}_{14}(t)&=&-2i\omega\rho_{14}(t)-2\lambda Y(t)\rho_{14}(t),\\ \non
\dot{\rho}_{22}(t)&=&2\lambda\Re[Y(t)]\rho_{11}(t)-2\lambda\Re[X(t)]\rho_{22}(t)-\lambda
X^*(t)\rho_{23}(t)-\lambda X(t)\rho_{23}^*(t),\\ \non
\dot{\rho}_{23}(t)&=&2\lambda\Re[Y(t)]\rho_{11}(t)-2\lambda\Re[X(t)]\rho_{23}(t)-\lambda
X(t)\rho_{33}(t)-\lambda X^*(t)\rho_{22}^*(t),\\ \non
\dot{\rho}_{33}(t)&=&2\lambda\Re[Y(t)]\rho_{11}(t)-2\lambda\Re[X(t)]\rho_{33}(t)-\lambda
X^*(t)\rho_{23}^*(t)-\lambda X(t)\rho_{23}(t),\\ \non
\dot{\rho}_{24}(t)&=&-i\omega\rho_{24}(t)+\lambda(Y(t)+X^*(t))(\rho_{12}(t)+\rho_{13}(t))-\lambda
X(t)(\rho_{24}(t)+\rho_{34}(t)),\\ \non
\dot{\rho}_{34}(t)&=&-i\omega\rho_{34}(t)+\lambda(Y(t)+X^*(t))(\rho_{12}(t)+\rho_{13}(t))-\lambda
X(t)(\rho_{34}(t)+\rho_{24}(t)).
\eeqa
Note the RDM elements are mutually coupled. $\rho_{11}$ will decay exponentially as $Y$ approach its steady value. If the initial state of the system does not contain $|11\rangle$ state, we can obtain the following analytical solution:
\beqa\label{analyall}
\rho_{22}(t)&=&\frac{1}{4}(1+|A|^2+2\Re[A])\rho_{22}(0)+\frac{1}{4}(1+|A|^2-2\Re[A])\rho_{33}(0)+\frac{1}{2}(|A|^2-1)R_{23}(0)+\Im[A]I_{23}(0), \\  \non
\rho_{33}(t)&=&\frac{1}{4}(1+|A|^2+2\Re[A])\rho_{33}(0)+\frac{1}{4}(1+|A|^2-2\Re[A])\rho_{22}(0)+\frac{1}{2}(|A|^2-1)R_{23}(0)-\Im[A]I_{23}(0), \\  \non
R_{23}(t)&=&\frac{1}{4}(|A|^2-1)\rho_{22}(0)+\frac{1}{4}(|A|^2-1)\rho_{33}(0)+\frac{1}{2}(|A|^2+1)R_{23}(0), \\  \non
I_{23}(t)&=&\frac{1}{2}\Im[A](\rho_{33}(0)-\rho_{22}(0))+\Re[A]I_{23}(0), \\  \non
\rho_{24}(t)&=&\frac{1}{2}e^{-i\omega
t}[A(\rho_{24}(0)+\rho_{34}(0))+\rho_{24}(0)-\rho_{34}(0)],\\  \non
\rho_{34}(t)&=&\frac{1}{2}e^{-i\omega
t}[A(\rho_{34}(0)+\rho_{24}(0))-\rho_{24}(0)+\rho_{34}(0)],\\  \non
A(t)&=&\frac{2\sqrt{\gamma}}{\beta}e^{-\frac{1}{2}\gamma t}e^{i\frac{1}{2}\Delta
t}\cos(\frac{1}{2}\beta t+\arctan\frac{i\Delta-\gamma}{\beta}).
\eeqa

\end{document}